# Investigating Socio-spatial Differences between Solo Ridehailing and Pooled Rides in Diverse Communities

**Jason Soria**
jason.soria@u.northwestern.edu

**Amanda Stathopoulos[1] (corresponding author)**
a-stathopoulos@northwestern.edu

## Abstract

Transformative mobility services present both considerable opportunities and challenges for urban mobility systems. Increasing attention is being paid to ridehailing platforms and connections between demand and continuous innovation in service features; one of these features is dynamic ride-pooling. To disentangle how ridehailing impacts existing transportation networks and its ability to support economic vitality and community livability it is essential to consider the distribution of demand across diverse communities. In this paper we expand the literature on ridehailing demand by exploring community variation and spatial dependence in ridehailing use. Specifically, we investigate the diffusion and role of solo requests versus ride-pooling to shed light on how different mobility services, with different environmental and accessibility implications, are used by diverse communities. This paper employs a Social Disadvantage Index, Transit Access Analysis, and a Spatial Durbin Model to investigate the influence of both local and spatial spillover effects on the demand for shared and solo ridehailing. The analysis of 127 million ridehailing rides, of which 15% are pooled, confirms the presence of spatial effects. Results indicate that density and vibrancy variables have analogue effects, both direct and indirect, on demand for solo vs pooled rides. Instead, our analysis reveals significant contrasting effects for socio-economic disadvantage, which is positively correlated with ride-pooling and negatively with solo rides. Additionally, we find that higher rail transit access is associated with higher demand for both solo and pooled ridehailing along with substantial spatial spillovers. We discuss implications for policy, operations and research related to the novel insight on how pooled ridesourcing relate to geography, living conditions, and transit interactions.

Keywords: spatial Durbin model, ridehailing, ride-pooling, pooled rides, spatial spillover, social disadvantage index

# 1. Introduction

A growing portfolio of urban mobility services are offered by Transportation Network Companies (TNCs) around the world. The new generation of on-demand and shared service models are poised to alter how cities fulfill their mission to provide citizens access to goods, services, and opportunities (Shaheen & Cohen, 2018). Ridehailing (RH) promises to offer more options to urban travelers, improve access to transit by providing first-last mile connections, increase vehicle occupancy via pooling, and offer on-demand flexibility for customers (Alonso-Mora et al., 2017; Shaheen & Cohen, 2018). However, the promise of RH has been questioned in recent studies (Diao et al., 2021). Empirical studies have shown that RH tends to be used for recreational trips rather than transit last mile access, and leans towards substitution effects with transit (Alemi et al., 2018; Tirachini & del Rio 2019). Additionally, several researchers find that surveyed RH users are likely substituting active modes like walking and biking (Clewlow & Mishra, 2017; Rayle et al., 2016), and that RH can generate induced demand (Rayle et al 2016, Tirachini & Gomez Lobo, 2019).

Research findings are also evolving to account for the constant service evolution of RH. The creation of shared RH service alternatives (also known as ride-pooling), such as UberPool, Lyft Line and Didi ExpressPool, match ride requests and give users a discount relative to the standard trip fare. These trips are authorized to be pooled and may possibly only serve one party when the demand is too low to efficiently match rides. In this paper we will refer to this service as ride-pooling, pooling, or pooled rides. We will also refer to the standard service (e.g. UberX and Lyft Classic) as solo rides since this service is exclusive to one party (a party may consist of more than one rider). Though RH has introduced a relatively more affordable alternative, RH in general is mainly used by narrow population segments, and consistently producing low shares of ride-pooling (Lewis & MacKenzie, 2017; Li et al., 2019; Rayle et al., 2016).

To date there is limited understanding of how RH demand is shaped by different community contexts and the degree to which solo and pooled differs from pooling (Soria et al., 2020). More commonly, these modes are not differentiated. In this paper we expand the literature on RH demand by using spatial modelling to examine the socio-economic community determinants. Specifically, we compare solo and pooled trip-making patterns from a large-scale Chicago database to identify the unique determinants that encourage pooled rides while controlling for spatial effects. The results of our analysis uncover new insights on how RH ties in with community factors, the importance of accounting for spatial effects, and whether solo and pooled rides serve distinct communities.

## 1.1. Ridehailing demand: user, context and community determinants

Analysis of the sociodemographic profile shows that RH adoption is higher in population segments that are younger, wealthier, more educated, and from smaller households (Alemi et al., 2018; Clewlow & Mishra, 2017; Dias et al., 2017; Rayle et al., 2016; Wang et al., 2019; Yu & Peng, 2019). It is also related to lower car ownership and to vehicle disposal (Alemi et al., 2019; Gehrke et al., 2019; Lavieri & Bhat, 2019a). Considering interactions with other modes, most RH trips seem to substitute the use of taxi or transit. Yet the precise demand relationship of RH with more sustainable options like public transit, walking and cycling is still unclear (Ward et al., 2019). Survey research suggests RH is a substitute for a sizeable share of users: Clewlow and Mishra (2017) suggest 15% of RH users would have used train, Feigon and Murphy (2016) indicate that 15% would have used bus *or* train, and 17% would have used active modes. Moreover, there is evidence that the availability of RH induces trips. Researchers find that 8% to 22% of trips would not have been taken if RH were not available (Rayle et al. 2016; Clewlow & Mishra, 2017; Henao, 2017). To date, little is known about the spatial, and population segment differences in substitution. Via an intercept survey, Gehrke et al. (2019) suggest that hailing-to-transit substitution is higher when general transit service is poor or unavailable, or RH costs less than 20$ (suggesting a shorter trip). Unexpectedly, however, lower income households were *more* likely to substitute transit for comparatively costly RH. This finding implies an equity concern where RH could fill mobility gaps for carless households with poor transit accessibility, while also straining those households' budgets.

Recent studies have modelled RH demand from publicly available trip data matched with land-use and socio-demographics. Results suggest that RH use is higher in areas with high population density, a higher proportion of high-income residents (especially for weekday trips), a higher proportion of younger residents, and more land-use diversity (Brown, 2019; Ghaffar et al., 2020; Lavieri et al., 2018; Marquet, 2020; Yu & Peng, 2019). These studies confirm survey-based research on user-profiles.

Findings related to the social context are less clear. An Austin study controls for ethnic composition in zones and find that RH demand is higher in areas with a lower share of white residents (Lavieri et al., 2018). Work from Boston indicates that RH users reflect the ethnic makeup of the area (Gehrke et al., 2019). These findings contrast with other research finding that new mobility platforms are used less in areas with a higher share of Latino and black residents (Biehl et al., 2018; Dias et al., 2019). Rather than taking these findings to suggest different racial and ethnic groups have intrinsic differences in their willingness to use RH, there are likely more nuanced correlation across income, social conditions, transportation and residential location patterns explaining these differences.

### 1.2. Ridehailing relationship with transit

Research on the relationship between RH and public transit warrants more attention as findings are mixed. Early work suggested complementarity as RH can fill transit accessibility gaps (Shaheen & Chan, 2016) and be accessed via public-private partnerships (Feigon & Murphy, 2016). Recent empirical studies from large-scale data tends to find that public transport use is positively correlated with ride-hailing use. Importantly, this is not in itself sufficient to conclude that a complementarity effect is dominant, since the aggregate correlational analysis cannot prove the exact impact of RH on public transport demand. Moreover, the precise measurement of transit access, as well as differences across contexts, and different service offerings (both bus vs rail and solo RH vs pooling) have not been clearly established. Several metrics of transit service have been utilized to research the connection between public transit and RH demand. Transit access time (TAT, i.e. the time to access a mass transit station), shows that poor subway access is related to fewer RH trips (Correa et al., 2017). Brown (2019) finds a similar relationship with higher overall transit stop density being associated with more TNC trips. Using National Household Travel Survey (NHTS) data, (Mitra et al., 2019) use a binary variable to denote rail service presence in the local statistical area, showing a positive effect on TNC trip demand for seniors. And finally, as the number of bus service hours increases, RH use also increases (Yan et al., 2020). Taken together, the research indicates a positive demand relationship between ride-hailing and public transit. We note however that other research suggests demand is traded off, or varied. In Lavieri et al. (2018), higher bus frequency is associated with less RH. A more nuanced, inverted U relationship is suggested in Ghaffar et al. (2020). That is, TNC ridership is lower in areas with the lowest and highest amount of bus stops, whereas a moderate number of stops boosts RH demand. Kong et al. (2020) study DiDi trips, and find trade-off effects in dense transit-rich areas (bus and subway), while suburban areas exhibit more complementary effect between transit and RH. Beyond the direct availability metrics, transit access to jobs is found to have a positive effect on TNC usage (Yu & Peng, 2019). And finally, Grahn et al. (2020) does not definitively conclude any relationship between RH and transit since transit includes a wide variety of modes (e.g. buses, light rail, commuter rail), each with different interactions with RH.

### 1.3. The potential for ride-pooling

Ride-pooling has the potential to reduce the number of passenger vehicles or vehicle miles travelled on the road assuming riders substitute personal or solo vehicle travel when opting to share. Simulation work suggests TNC fleet sizes can be reduced with shared rides (Alonso-Mora et al., 2017). However, the share of pooling likely needs to be much higher than currently observed to unlock benefits. Rodier et al. (2016) suggests above 50%, while Fagnant and Kockelman (2018) estimate that pooled services need to account for 20–50% of the market-share. To date, little is known about the current demand for pooled rides nor the determinants of use. Basic statistics are uncertain but suggest a market-share of pooling between 6 and 35 % (California Air Resource Board, 2019; Chen et al., 2018; Chicago Metropolitan Agency for Planning, 2019; Li et al., 2019; Lyft, 2018; Soria et al., 2020; Young et al., 2020). The hypothetical demand for pooling has been examined in stated preference work, finding that the addition of co-riders generates non-linear disutility in a shuttle setting and high sensitivity to time-cost trade-offs (Alonso-González et al., 2020). In the context of a shared autonomous rides, Lavieri and Bhat (2019b) also suggest that the travel time/waiting time to cost trade-offs matter more than the perceived disutility of sharing a ride. Sarriera et al (2017) also find that time and cost considerations outweigh social interaction effects. In terms of mode-substitution, survey data from Hangzhou, China suggests that the biggest mode-shift of ride-pooling users would be to transit (bus and metro rail) (Chen et al., 2018).

Recently, a limited number of major RH data-releases is supporting initial empirical analysis of pooling. Analysis of large-scale trip data suggests that solo and pooled demand has different spatio-temporal patterns in Chengdu, China (Li et al., 2019). Ensemble machine learning highlights the importance of pricing and timing variables for ride-pooling demand in Hangzhou, China (Chen et al., 2017). Clustering analysis on Chicago RH data reveals that pooled rides have distinct patterns, linked to affordability and local transit performance (Soria et al., 2020). These works shed light on the user trade-offs and aggregate demand patterns of ride-pooling. Yet we still know little about the hurdles to the increased adoption of pooled rides to reach the critical mass needed to unlock significant mobility benefits in terms of VMT reductions.

### 1.4 Spatial modelling of mobility impacts

There is ample evidence that transportation infrastructure is often associated with "broader" impacts via analysis of surrounding or neighboring spatial units (e.g. states, counties, Census Tracts). Yu et al. (2013) find that transport infrastructure capital (roadways, railways, water transport, and civil aviation) in China has a positive spillover effect on GDP across regions; Berechman et al. (2006) find strong spillover effects of highway capital investment in the US. Additionally, urban rail projects in the US have been tied to increased residential property values in surrounding areas (Chen et al., 1998; Diao, 2015). Similarly, other spillover effects such as increases in household income have been observed around urban rail stations in Denver, CO (Bardaka et al., 2018). Not all spillovers are positive, though.  observe negative spillovers of nuisances such as noise associated with light rail transit. In practice, investments such as light rail construction, often comport both positive (accessibility) and negative (nuisance) effects spillover effects (Chen et al 1998). In addition, the spatial distribution of new transportation infrastructure is often distributed unevenly with regard to race and socioeconomic status of residents. Hirsch et al (2017) found that health-promoting infrastructure (parks, bicycle facilities, off-road trails, and public transportation) in four US cities was spatially clustered, and often associated with income and employment status of residents. In sum, spatial spillovers exist, and often play an important role in terms of equity and health disparities. Knowing the nature and degree of spillovers related to transportation investments has evident practical value by improving planning and accounting for the equity in distribution of spillover effects across areas (Cohen, 2010).

Little is known on the potential spatial aspects of RH operations. This analysis is complicated by the spatio-temporal variation in on-demand services, limited data on both demand and supply, as well as continuous regulatory and service evolution. Research by Hughes and MacKenzie (2016) compared spatial variability in wait times for UberX throughout the Seattle region. Wait times increased in areas with higher average income and decreased in areas with greater population and employment density. Brown (2018) directly compared Lyft and taxi performance for Los Angeles, California. She observed that RH serves more diverse neighborhoods and have lower cancellation rates and waiting times than traditional taxis. Other studies examine the competition between taxis and RH by accounting for spatial differences. Kim et al. (2018) study the spatial effects TNCs have on New York City taxis where RH's entry decreased taxi demand in one part of the city while increasing it in others. In other markets, RH filled spatial and temporal gaps in taxi supply (Dong et al., 2018). Moreover, initial evidence from observed trip-data suggests robust spatial differences between solo and pooled rides (Chen et al 2017; Li et al., 2019; Soria et al., 2020). With limited analysis it is difficult to draw general conclusions about spatial variation in the demand and impact of RH, though we note that the effects appear to be dynamic and tied to local community conditions. A deeper analysis of different spatial patterns that also account for socio-economic conditions and land-use variables, is needed to understand ride-pooling and inform better policies to maximize their benefits for users across diverse urban environments.

### 1.5. Literature gaps and research motivation

On the whole, the diffusion of RH appears to be related to existing socio-economic and mobility advantage of users. Despite the significant growth in use, suggesting that 36% of U.S. adults have now tried RH (Pew Research Center, 2018), adoption disparities persist, most notably between urban and rural communities, younger and older users, and income groups (Alemi et al., 2018; Alonso-González et al., 2020; Lavieri & Bhat, 2019a). While the adoption gaps among population segments is well established, the spatial gaps in use and service, including relationships to competing modes, are still unclear.

For both general RH and ride-pooling analysis, most previous work typically uses an "aspatial" perspective, explaining usage patterns by accounting for characteristics within the spatial unit of analysis, but not controlling for spatial correlations nor investigating spillovers across neighborhoods. Ghaffar et al. (2020) and Dean and Kockelman (2021) consider similar socio-economic, built environment, and transit accessibility variables with methods that consider spatial effects with Chicago RH data. These studies use census tracts as the spatial unit of investigation. This research considers Chicago Community Areas as the spatial unit of investigation to include approximately 24% of the data that are missing due to trip origin censoring.

The definition of Chicago Community Areas and information about trip origin censoring are provided in the Methods and Materials section.

### 1.6 Research objectives

We complement the existing research that considers spatial effects by considering a Spatial Durbin Model (SDM) (Dean & Kockelman, 2021; Ghaffar et al., 2020; Lavieri et al., 2018; Yu & Peng, 2019). Additionally, we investigate and compare determinants of demand for solo and pooled ride demand in depth. Previous research does consider ride-pooling separately and finds that it is different from solo rides based on average travel time and distance, time of day when it is most utilized, and general economic indicators such as gross domestic product and average house price (Li et al., 2019). To build upon this research, we account for socio-economic, land-use, and rail access time variables to understand community dynamics of RH adoption, including community level spillovers. Methodologically, we employ the SDM (Anselin, 2003). This approach enables us to investigate whether the intensity of RH demand in a community area is associated with the features of the observed area, as well as of its neighbors. In this paper we focus on three research objectives that each make a contribution to understanding RH demand determinants.

- *Q1:* What are the *spatial patterns of demand for solo and pooled rides, and do they differ?* This research contributes to building fundamental insight from large-scale data on pooled demand distinctions. We further explain differences in Q2 and Q3.
- *Q2: What is the impact of socio-economic conditions of communities on RH demand (solo and pooled)?* The specific contribution is to account for the bundled nature of socio-spatial advantage/disadvantage indicators and provide new insight on how pooling and solo RH relates to community disadvantage.
- *Q3: What is the demand-relationship between RH (solo and pooled) and transit accessibility?* This research contributes to more understanding of the still mixed findings of how RH relates to transit.

Our findings from the SDM analysis of Chicago RH demand coupled with auxiliary data suggests uniformity in effects for land-use and density variables. Instead, solo and pooled demand has nuanced and diverse effects when considering transit competition and social disadvantage impacts.

## 2. Materials and Methods

### 2.1 Ridehailing data and dependent variable definition

RH demand data (plus metadata for spatial boundaries) are collected from the City of Chicago public data portal (City of Chicago, 2020). The dataset details all TNC trips within the city limits and records the origin and destination census tract, time of departure and arrival, total fare paid, and whether the trip was authorized to be shared (and if so, how many parties joined the trip). The data are processed and cleaned by removing observations with no origin or destination, fares of $0 and extremely high values (greater than $1,000), or 0 trip duration or miles recorded. The clean dataset used in this analysis comprises 127,598,605 ride records between November 2018 and December 2019. Approximately 25.5% of these trips were authorized to be shared, however, only 66.9% of these were truly shared, indicating that overall, 17% of all trips were truly pooled.

To preserve privacy, the Census tract info is censored if only one trip occurs in a 15-minute interval, and spatially aggregated up to the community area level. These types of trips account for nearly 24% of the data. Owing to this restriction, and the availability of auxiliary data, we opt to model RH demand at the more aggregate spatial level of the 77 community areas defined by the city. These community areas were originally based on groups of neighborhoods and physical barriers (Owens, 2012). Using this spatial unit of analysis is also advantageous because the boundaries rarely change, unlike Census based spatial units.

The trip data were aggregated based on trip origins which are the most likely to reflect the socio-economic origin of users, though we note that destination, or OD pairs could be used (Ni et al., 2018). We assume that the attributes of trip origins are reasonably good descriptors of riders' demographics with Young et al. (2020) finding that 86.4% of trips were home-based. Owing to the varying sizes of community areas, the ridership data are normalized by the area of the communities (in square miles). Additionally, trip demand is heavily skewed towards the downtown areas. To account for this, a log transformation is applied. The dependent variable thereby represents long-term RH intensity while controlling for community area and demand intensity variation.

## 2.2 Transit access measure

Studies have found that transit can play either a competing or complementary role with no consensus on which relationship is stronger (Babar & Burtch, 2017; Boisjoly et al., 2018; Hall et al., 2018; Nelson & Sadowsky, 2019; Young et al., 2020). To add to this discourse, we model the impact of transit accessibility on the intensity of RH usage. The location of all Chicago Transit Authority and Metra public transit rail stations are collected from the public data portal (City of Chicago, 2020). The transit accessibility measure used in this study is akin to the Transit Access Time defined by Correa et al. (2017) where a hexagonal tessellation is overlaid on a map of the city. The edge of each cell is 1750 ft so that the theoretical walking time across is within the pedestrian access time defined by the Federal Highway Administration guidelines (Nabors et al., 2008). To determine average transit access time in each community area, the Google Maps API is used to determine the walking time from the center of a hexagon to the closest rail transit stop and averaged across the community area (Google, 2020). A similar approach was used to derive bus station density, but this measure was found to be insignificant in modeling.

## 2.3 Social Disadvantage Index

Across cities, urban mobility systems naturally intersect with long-running challenges, including spatial mismatch, enduring racial residential segregation and economic inequality. For Chicago, it is known that economically depressed areas tend to be poorly served by transit (The Chicago Urban League, 2016). The local planning agency, CMAP has called for more research to examine the benefits and pitfalls of new mobility technologies, such as RH, with regard to accessibility, affordable mobility, and quality of life in underserved communities (CMAP, 2018).

Moreover, work in the social sciences has established that numerous factors related to household structure, employment, income, wealth and racial status can make households more vulnerable to a lack of economic opportunity that is perpetuated as economic immobility (Sabol et al., 2020). Moreover, just like socio-demographic privilege, disadvantage comes in clusters, making it difficult to allocate the influence of separate factors (Smeeding, 2016). To date, existing RH research has limited analysis to single socio-demographic factors, like race or income. In this paper we parse the simultaneous dimensions of socially disadvantaged communities and how they correlate with the adoption of RH services by developing a Social Disadvantage Index (SDI). A similar index has been applied to examine the relationship between measures of deprivation and health outcomes (Butler et al., 2013). To determine the SDI we rely on data from the ACS 5-year estimates for the Census tracts, aggregated to the Chicago community area level (U.S. Census Bureau, 2019). A single factor Exploratory Factor Analysis (EFA) with a factor loading threshold of 0.30 and no rotations is used to obtain an SDI for each community area. The composition of the SDI is summarized in Table 1.

The index has intuitive results and high internal validity (Cronbach's $\alpha$ =0.91), suggesting strong links between household income and a number of vulnerability factors. The advantage of using an index is to enable a more holistic analysis that does not define hardship by looking at single factors such as racial or ethnic minority status. Instead, the validity of the proposed factor analysis affirms the strong correlations among disadvantage metrics, and the risk of spurious results should the items be included separately.

**Table 1 Social Disadvantage Index Results**

| Item | Factor Loadings |
| --- | --- |
| Percent of population with poverty level income | 0.989 |
| Percent of households with single parent | 0.873 |
| Percent of population that are non-white | 0.769 |
| Percent of households with no vehicle | 0.763 |
| Percent of households renting for housing | 0.744 |
| Percent of working eligible that are unemployed | 0.649 |
| Cronbach's $\alpha$ | 0.91 |

\* Result from Exploratory Factor Analysis on ACS data, unrotated single-factor results

### 2.4 Land-use and demographic variables

Beyond the SDI that captures economic vulnerability, our analysis controls for other relevant socio-demographics that have been tied to RH demand in the literature: user age, household size, and population density (Clewlow & Mishra, 2017; Lavieri & Bhat, 2019a; Rayle et al., 2016). We collected this data from the ACS (U.S. Census Bureau, 2019).

RH use is also associated with land-use mix (Ghaffar et al 2020). We define a land-use mix index, following Ghaffar et al. (2020), and measure it at the community area level using data from CMAP (CMAP, 2018). This index was tested in our model specification but did not yield statistically significant results. Given the connection of RH use to recreational and leisure travel (Soria et al 2020), we extract data on the location of restaurants and bars with active licenses during 2018 and 2019 (City of Chicago, 2020). This measure represents the impact of *third places*, namely the localities that are separate from home and work that generates a sense of community and contributes to urban vibrancy (Oldenburg & Brissett, 1982; Trentelman, 2009). The bar/restaurant variable is normalized by area.

Table 2 shows key socio-demographics, transit access, and RH characteristics of major Chicago Districts (collection of community areas). We note that the areas with higher income (North and Central) tend to have better transit access (lower TAT) and more RH pickups, but lower degrees of pooling, albeit with some variation across communities. Figure 1 maps the delimitations of these districts. Table 3 shows the summary statistics for all dependent and independent variables included in the final models. It shows that the model includes highly diverse communities with wide ranges of youth population, population density, bar and restaurant density, and TAT.

Importantly, because the factor analysis only includes ACS data from Chicago, the SDI cannot be directly compared with other cities.

**Table 2 Descriptive Statistics of 9 Major Chicago Districts**

| | Chicago | Far North | Far NW | North | Central | West | South | SW Side | Far SE | Far SW |
|---|---|---|---|---|---|---|---|---|---|---|
| Avg. Income Per Capita ($) | 32,535 | 33,744 | 25,172 | 57,393 | 87,061 | 26,755 | 24,364 | 17,570 | 19,737 | 26,682 |
| Avg. Income Per Household ($) | 84,637 | 82,208 | 76,569 | 126,994 | 147,138 | 76,703 | 56,731 | 58,606 | 54,302 | 77,102 |
| HS Degree only (% of pop) | 23% | 20% | 28% | 11% | 6% | 24% | 23% | 37% | 30% | 27% |
| Bachelor's or higher (%) | 36% | 45% | 24% | 66% | 73% | 31% | 30% | 11% | 18% | 26% |
| Commuting SOV (%) | 53% | 53% | 67% | 41% | 32% | 49% | 46% | 62% | 63% | 71% |
| Commuting Carpool (%) | 8% | 7% | 11% | 5% | 4% | 9% | 8% | 14% | 9% | 9% |
| Commuting Transit (%) | 30% | 32% | 19% | 45% | 30% | 31% | 33% | 20% | 26% | 19% |
| Commuting Active (%) | 9% | 7% | 4% | 9% | 34% | 11% | 12% | 4% | 3% | 1% |
| Avg. Rail Access Time (min) | 24.1 | 38.9 | 21.9 | 13.3 | 11.0 | 14.7 | 12.5 | 24.7 | 31.0 | 22.1 |
| Avg Daily TNC Pickups | 263,192 | 27,030 | 7,763 | 53,727 | 77,952 | 57,390 | 18,833 | 11,371 | 5,619 | 3,507 |
| Avg Daily Authorized Shared TNC Pickups | 59,006 | 6,644 | 2,439 | 8,868 | 11,261 | 14,177 | 6,879 | 4,614 | 2,575 | 1,549 |
| TNC Rides Authorized to be Pooled (%) | 22% | 19% | 28% | 17% | 14% | 24% | 37% | 35% | 41% | 37% |
| TNC Rides Truly Shared (%) | 15% | 12% | 18% | 12% | 11% | 18% | 26% | 23% | 23% | 22% |
| Share of Authorized Pooled Rides that are truly shared (%) | 69% | 65% | 64% | 73% | 77% | 72% | 70% | 65% | 57% | 58% |

**Table 3 Model Variable Summary Statistics**

| Variable | Median | Mean | Standard Deviation |
|---|---|---|---|
| Dependent Variable: Log of Average Daily Solo Trips per square mile | 5.698 | 5.957 | 1.357 |
| Dependent Variable: Log of Average Daily Shared Trips per square mile | 5.316 | 5.196 | 1.119 |
| Population 18yr to 34yr (%) | 0.2530 | 0.2735 | 0.07741 |
| Population Density (per sq. mile) | 11,521 | 13,113 | 7,002 |
| Mean Household Size | 2.716 | 2.739 | 0.5407 |
| Bar and Restaurant Density (per sq. mile) | 35.559 | 58.058 | 73.76 |
| Transit Access Time (minutes) | 14.569 | 19.756 | 13.45 |
| Social Disadvantage Index | -0.1683 | 0 | 0.9904 |

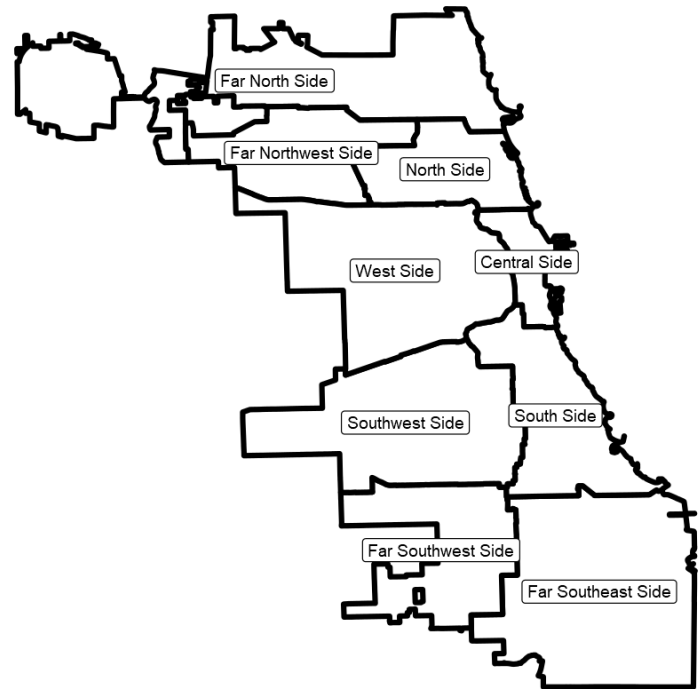

Figure 1 Chicago Area District Map
*Note. Bold borders depicting the boundaries of the Chicago sides*

## 2.5 Methodology: Spatial Durbin Model

Previous transportation research investigating RH use has relied on representation of the context measuring only the "immediate spatial area", with limited investigation of factors occurring in surrounding areas. Importantly, while a portion of the impact is determined in the immediate spatial area, some effects are likely to spill over across communities. This spill-over is not directly tied to demand awareness. Instead, while it is unlikely that riders are directly aware of RH demand in neighboring areas, the local and surrounding community conditions are likely to affect demand for RH via waiting times and social effects. That is, local mobility praxis, driver pickup biases and strategies, and perceived attractiveness and viability of alternatives can all shape spatial (spillover) demand for RH. To study this, we regress the intensity of both solo and pooled usage on a range of potential explanatory factors. We find evidence for a significant role of transit accessibility, SDI, along with four land-use/density variables, summarized in Table 3.

We apply spatial econometrics to account for spatial interactions (Manski 1993). After verifying the presence of spatial autocorrelation, and using Moran's I and Lagrange Multiplier tests for model specification guidance, we specify a Spatial Durbin Model (SDM) to explore our three research questions (Anselin & Kelejian, 1997; Osland, 2010). The general SDM specification is summarized in equation 1. Y is the response variable of community area RH demand, $\rho$ is a coefficient for the lagged effect representing the response variable in one community to other neighboring communities, and W is a weight matrix representing the spatial structure of community influences on the residuals. This first term, $\rho WY$, measures the endogenous effect of RH usage. The spatial weight matrix, W, is defined as a row-standardized matrix where each row represents the spatial unit of analysis, contiguous neighbors have an equal effect, with 0's along the diagonal. The row sum of the weights is equal to 1 for every spatial unit. The purpose of using the row-standardized weight matrix is two-fold. First, a row standardized matrix facilitates efficient maximum likelihood estimation of the SDM (LeSage & Pace, 2009). Secondly, the row normalization of W means that the effect of neighbors is averaged which is desirable when there is no a priori knowledge of neighbor influence. This W is used throughout the modeling to maintain comparability. X is a matrix of explanatory variables and $\beta$ is the vector of corresponding coefficients. $\gamma_l$ is the vector of spatial lag coefficients of the explanatory variables $X_l$. An extension of this model is the Spatial Durbin Error Model (SDEM) which considers the error term as a function of W.

Because SDM includes an endogenous term, the estimated coefficients are not representative of the impacts of the explanatory variables. To translate them into interpretable values, the coefficients are transformed. Equations 2, 3, and 4 are used to obtain direct (immediate local effects), indirect (spillovers), and total impacts (the sum), respectively, to examine the impacts of the explanatory factors on both solo and pooled ridehailing. These impacts are calculated for each explanatory variable, k, using the $\rho$ estimated in equation 1.

$$Y = \rho WY + X\beta + WX_l\gamma_l + \epsilon \tag{1}$$

$$Direct = \frac{3 - \rho^2}{3(1 - \rho^2)}\beta_k + \frac{2\rho}{3(1 - \rho^2)}\gamma_k \tag{2}$$

$$Indirect = \frac{3\rho + \rho^2}{3(1 - \rho^2)}\beta_k + \frac{3 + \rho}{3(1 - \rho^2)}\gamma_k \tag{3}$$

$$Total = \frac{3 + 3\rho}{3(1 - \rho^2)}(\beta_k + \gamma_k) \tag{4}$$

## 3. Results and Discussion

### 3.1 Mapping of ridehailing variables

Before analyzing the model results, we explore the general patterns of demand for RH along with ACS data. Figure 2 depicts the percent of RH rides that are solo (a) and pooled (b), respectively. We also plot the SDI scores by community area in Figure 3. Comparing Figures 2 and 3 suggests the community areas with higher SDI index (more disadvantaged) tend to rely more on ride-pooling, as these maps have stronger spatial similarity. The trends are most evident with central and northern communities exhibiting lower rates of sharing and low SDI whereas western and southern community areas have higher rates of sharing with a higher SDI. Along with the statistics on ride-pooling shown in Table 2, this provides initial evidence that the spatial dynamics of solo and pooled rides differ and have strong ties to socio-economic vulnerability.

### 3.2 Spatial Durbin Model specification

Given the strong differences in spatial patterns of solo and pooled rides, we estimate separate models. The modeling starts with a bottom-up approach: estimating non-spatial linear regression models by OLS including all the theorized RH demand drivers. Residual diagnostics and the Moran's I-test is used to detect spatial dependency. Both solo rides (Moran's I = 0.30705, p-value = 0.001) and pooling (Moran's I = 0.37534, p-value = 0.001) gives evidence of spatial autocorrelation. Thereby we follow Elhorst's (2010) combined approach using Lagrange multiplier (LM) and likelihood ratio testing. With the need to control for spatial effects apparent, the Lagrange Multiplier (LM) test is used to determine the need for spatial lag or spatial error controls. The spatial lag (statistic = 34.35, p-value < 0.001) and spatial error (statistic = 17.03, p-value < 0.001) model specifications indicate that either approach is potentially valid. However, with both tests significant, the SDM is favored over a potential SDEM because it is more robust (Osland, 2010). The estimation of the SDM was completed using the R programming language and spatialreg package (Bivand & Piras, 2015; R Development Core Team, 2008). Further comparison of SDM and SDEM likelihood ratio tests and inspection of spatial correlation confirms that the former provides more interpretable findings. **Table 4** and **Table 5** show the regression results and impacts, respectively. The following section discusses the interpretation of the findings followed by a deeper analysis of the three research questions.

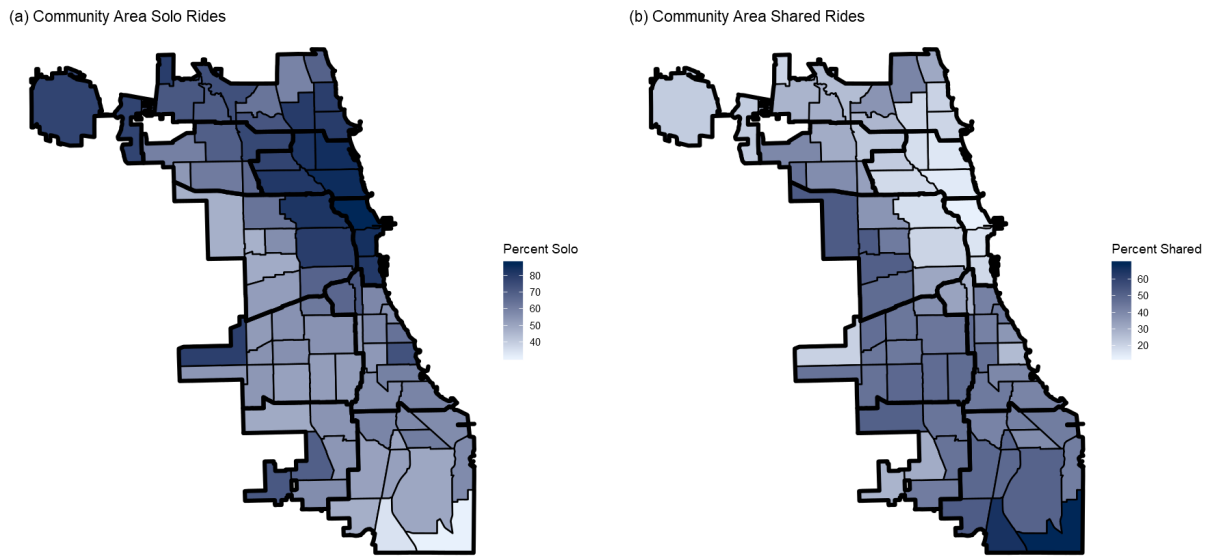

Figure 2 Community Area Percent use of Solo (a) and Ride-pooling (b) Map with bold borders depicting the boundaries of the Chicago sides

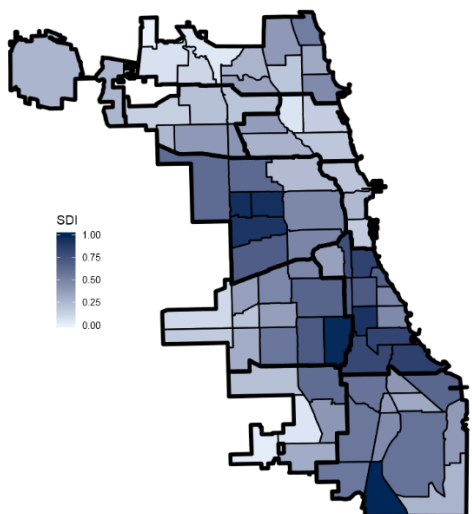

Figure 3 Social Disadvantage Index mapped by community area

**Table 4 Spatial Durbin Model Estimation Result[1]**

| Variable | Solo Rides | | Authorized Ride-pooling Rides | |
|---|---|---|---|---|
| | Coefficient | t-Statistic | Coefficient | t-Statistic |
| (Intercept) | 3.90*** | 5.87 | 2.37*** | 4.3 |
| Population 18yr to 34yr (%) | 4.03*** | 4.06 | 2.68** | 3.26 |
| Population Density (100,000s per sq. mile) | 3.57** | 3.28 | 3.02*** | 3.47 |
| Mean Household Size | -0.387*** | -3.78 | -0.163* | -1.98 |
| Bar/Restaurant Density (1,000s per sq. mile) | 1.89^ | 1.88 | 1.17^ | 1.43 |
| Transit Access Time (minutes) | 0.00122 | 0.278 | -0.00904* | -2.48 |
| Social Disadvantage Index (score) | -0.124* | -2.38 | 0.146*** | 3.39 |
| Lag ($\gamma$) for Transit Access Time (minutes) | -0.0411*** | -4.32 | -0.0201* | -2.52 |
| $\rho$ | 0.369*** | | 0.508*** | |
| Nagelkerke Pseudo $\rho^2$ | 0.919 | | 0.917 | |
| AIC (OLS) | 90.985 (105.97) | | 63.232 (91.301) | |
| Residual Autocorrelation | 1.17 | | 4.41* | |
| n. community areas | 77 | | 77 | |

[1] - Several variables were tested and if found to be insignificant in both the Solo and Authorized Pooled models were removed from the model specification. These were: bus stop density, percent of land area dedicated to parks, and mixed land-use

^ - p-value < 0.1; * p-value < 0.05; ** p-value <0.01; *** p-value <0.001

**Table 5 Impacts of Explanatory Variables**

| | Solo Rides | | | Authorized Ride-pooling Rides | | |
|---|---|---|---|---|---|---|
| | Direct Impact | Indirect Impact | Total Impact | Direct Impact | Indirect Impact | Total Impact |
| Population 18yr to 34yr (%) | 4.17 | 2.23** | 6.40*** | 2.870 | 2.59** | 5.457*** |
| Population Density (100,000s per sq. mile) | 3.69*** | 1.97** | 5.66*** | 3.23*** | 2.91** | 6.14*** |
| Mean Household Size | -0.400^ | -0.214** | -0.614*** | -0.175^ | -0.157^ | -0.332* |
| Bar and Restaurant Density (1,000s per sq. mile) | 1.95** | 1.04 | 2.98^ | 1.25*** | 1.13 | 2.38 |
| Transit Access Time (minutes) | -0.00237 | -0.0608*** | -0.0632*** | -0.0124** | -0.0468*** | -0.0592*** |
| Social Disadvantage Index | -0.128* | -0.0685* | -0.196* | 0.157* | 0.141** | 0.297*** |

^ - p-value < 0.1; * p-value < 0.05; ** p-value <0.01; *** p-value <0.001

### 3.3. Direct and indirect effects on ridehailing demand

**Table 4** shows the SDM results with a spatial lag effect $\rho$ evident for both solo and pooled rides. The lagged $\gamma$ coefficient for TAT is highly significant (p-value < 0.001). This suggests that in both RH cases there is a need to account for spatial effects, including indirect impacts, most evident for transit accessibility. Both the solo and pooled ride demand models produce a high goodness of fit with Nagelkerke pseudo $\rho^2$ (similar to $r^2$ in OLS) greater than 0.90 and AIC lower than equivalent OLS specifications, all suggesting the SDMs are valid and justified. There is evidence of residual spatial autocorrelation in the ride-pooling model, but the significance is low. Because these coefficients are not directly interpretable, the impacts of explanatory variables are calculated via the spatialreg package in R and summarized in **Table 5** in the form of direct, indirect and total effects as described in equation 2-4 (Bivand & Piras, 2015; R Development Core Team, 2008). That is, a change in the independent variables in a community area will not only lead to a change in the demand in the same community (direct effect), but also affect the RH demand in other community areas (indirect effects, related to the off-diagonal elements in W).

There is evidence of six factors affecting the community area demand for RH with some variability in terms of direct and indirect impacts. To gain more intuitive understanding of the effects, we use equation 5 to compute impact measures, where $\Delta_x$ is the change in variable x and $I_x$ is the impact of variable x from **Table 5.** We thereby estimate changes in average daily requested rides. Interpreting the direct effects of population density, we find that an increase of 1000 in population density is associated with approximately 7,700 more solo rides and 2,000 additional pooled rides per day in that community. Using the average population density from **Table 3,** this translates to a 1% increase in population density being associated with a 0.49% and 0.42% increase in daily demand for solo and pooled rides, respectively. These findings do not account for the spillover effects into other community areas. Turning to investigate transit rail accessibility, given the pronounced indirect effects, the spillovers are computed instead. For example, if a rail station were removed and a community area's average rail access time increases by 1 minute, then the sum of changes in neighboring community areas results in 12,000 fewer solo rides and 2,700 fewer pooled rides. In terms of total (direct and indirect) impacts, on average, a 1% increase in TAT is associated with strong reduction in RH requests (-1.24% for solo; -1.16% for pooled).

$$\ln(r_2) - \ln(r_1) = \Delta_x I_x \ln\left(\frac{r_2}{r_1}\right) = \Delta_x I_x \qquad (5)$$

$$\frac{r_2}{r_1} = \exp\left(\Delta_x I_x\right)$$

$$r_2 = r_1 \exp\left(\Delta_x I_x\right)$$

$$r_2 - r_1 = \Delta_r = r_1(\exp(\Delta_x I_x) - 1)$$

In the following sections we turn to discuss the results in the context of addressing our three research questions.

### 3.4 Differences between solo ridehailing and ride-pooling

Our first goal is to investigate the spatial usage patterns of solo versus pooled ridehailing. Before studying the model results, we examine the spatial distribution of Community area centroid Origin-Destination flows. Comparing Figure 4 and Figure 5 reveals stark differences in the user patterns with a greater spatial dispersion of ride-pooling compared to highly concentrated OD flows of solo rides, illustrated by the red connectors concentrated in the downtown and airport corridors. Taken together, the mapping of RH intensity (Figure 2) and flows (Figures 4-5) strongly suggests that ridership patterns are distinct. We turn to the model results in Table 4 and 5 to formally examine the causes for these differences. Both solo and pooled RH demand is higher in community areas with higher population density, more bars and restaurants, and higher share of young (18-34yr) population, with slightly stronger impact of each factor for solo use. This leads to a first observation that urban vibrancy factors stimulate RH demand more broadly, with uniform impact on solo and pooled ride requests.

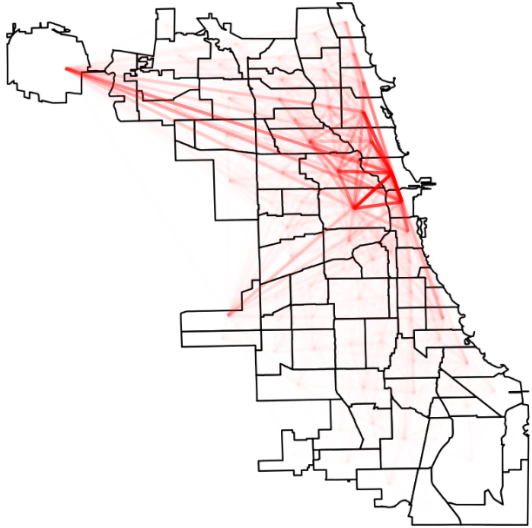 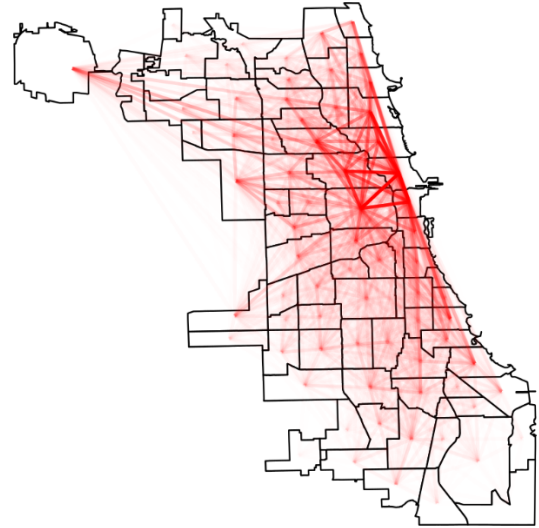

Figure 4 Intensity of OD flows of Solo Ridehailing

Figure 5 Intensity of OD flows of Ride-pooling

This allows us to confirm established research findings on the key role played by urban density variables, and to extend those findings also to pooled RH demand (e.g. Dias et al., 2019, Yu & Peng, 2019).

However, this leaves the question of explaining the prominent spatial differences for solo and pooled hailing open. The SDM models reveal that the main source of the divergent spatial patterns are the social disadvantage and transit accessibility metrics, examined further in Sec. 3.5 and 3.6. Notably, the relative socio-economic *disadvantage* of communities appears to be the main differentiator for pooled versus solo RH demand. This finding suggests an intriguing new connection between the evolving service portfolio of RH operators and diverse socio-economic demand segments. That is, the small share of dynamic ride-pooling requests are disproportionally requested in areas of socio-economic disadvantage, in contrast to the typically observed patterns of RH demand established in the literature.

### 3.5 Social Disadvantage Index and Spatial Effects

The second research question centers on exploring the socio-economic factors, and particularly the correlated socio-economic disadvantage observed across Chicago. The SDI analysis and mapping confirm the correlated nature, as well as the spatial concentration of socio-economic disadvantage indicators. The map in Figure 3, illustrates stronger vulnerability in the West, Southwest, South and Far Southeast districts (Fig. 1) of Chicago.

The modelling confirms that concentrated disadvantage is associated with fewer solo requests. This evidence supports the argument that (solo) ridehailing is related to ridership privilege (Lewis and MacKenzie, 2017). This is because RH, solo rides in particular, is offered as a premium service and at a higher price than other available mobility options in the area. Yet, this correlation with privilege is not supported by the model results for pooling requests. This suggests an intriguing interpretation that ride-pooling plays a gap-filling role for households with lower income and limited access to a personal vehicle (Tirachini and del Rio 2019). It is worth noting that this higher demand occurs despite the higher price-point of RH, even considering the discount for pooled rides. Considered jointly, we note that even after controlling for population and bar/restaurant density, the social and economic conditions in the community area still play an important role in shaping demand for RH. A crucial question that arises from these results is the contradiction of a higher number of requests for sharing, occurring in the very areas where demand is generally low and matching multiple trip trajectories is challenging. This is also reflected in the Table 2 statistics. The share of effectively matched trips (15%) is lower than the requested share of pooling (22%) with the rate of effective matching being highest in the wealthier central district and lowest in the far southeast district.

A second observation concerns the robust spillover effects for the SDI for shared rides. The negative indirect effect implies that disadvantage in adjacent community areas reinforces the direct demand effects. We attribute this indirect impact to the social nature of technology adoption (Alemi et al., 2018, Alemi et al., 2019). For other shared mobility services like bikesharing, there is evidence of social/community factors driving adoption (Manca et al 2019, Biehl et al 2019). It is not clear that RH embodies the same level of symbolism or spatial visibility that bikesharing does. Therefore, we propose as an area of future investigation to disentangle whether the observed spillover of socio-economic conditions is due to supply effects (RH drivers avoiding, or not opting in to offer pooled rides, in certain areas) or demand effects (local service/acceptability, social diffusion). For the latter case, we would specifically need to examine whether there are spatially bound social network effects leading to more use of pooled services, or whether the spatially correlated challenges of longer commutes and poorer mobility options (i.e. spatial mismatch) drive the needs for pooled RH to fill gaps in underserved community areas.

### 3.6 Spatial Effects and Rail Transit Access

The third research question probes the relationship between RH (solo and pooled) and local rail transit accessibility measured via the TAT variable. The research is still divided regarding the substitutional (Clewlow and Mishra 2017) or complementary (Boisjoly et al., 2018) relationship of RH with transit. Moreover, research suggests systematic variation is likely according to the size of the city (Hall et al., 2018), locations within a city (Grahn et al 2020), the number of TNC competitors in the market (Nelson and Sadowsky, 2019), and the transit option type and performance (Babar and Burtch, 2017). What is more, the existing research offers limited insight on the connection of pooled RH and transit.

Looking at Chicago, there are factors suggesting both relationships are possible. The strong variability in wealth and service access across the city could suggest *complementarity* since users, particularly in underserved community areas, may opt to use RH to fill gaps in transit accessibility (Alemi et al 2018), albeit with a need to consider the steep price differences (Hall et al., 2018). We would expect this to occur particularly for more affordable pooled rides. Instead, Chicago's expansive transit system with a high transit performance score (AllTransit 2020) suggests that transit could remain *competitive* even in the presence of multiple TNC operators, as suggested by Babar and Burtch (2017). Finally, the loop-centered radial nature of Chicago's CTA rail system points to possible variation in effects according to the north-south corridor.

On the whole, our SDM model results suggest a significant positive correlation between RH and rail accessibility. That is, in community areas where transit performs better (lower access times) the demand for RH is also *higher,* in line with Correa et al., (2017) and Brown (2019).

While this result is not surprising given the previous research using real trip data, the results are important because they can provide evidence of the separate effect of pooled rides. We expected that ride-pooling could have a more competitive demand relationship with rail transit, given the lower price-point and shared reliance on sharing. For example, Lewis and MacKenzie (2017) found that UberHOP, a ride-pooling service, predominantly drew riders form transit.

Instead, we find a significant direct demand effect only for pooled RH, and no significant differences overall between solo and pooled requests. Additionally, there is a strong spillover effects for TAT (Table 5), suggesting that transit accessibility in one community affects its neighbors. We attribute this to the spatial nature of transit systems where rail transit routes traverse several community areas.

In summary, in the central and northern areas of Chicago, excellent rail transit accessibility is correlated with higher demand for RH. A possible explanation is that RH competes more directly with driving than with transit, and the lower auto ownership and parking availability makes RH more attractive precisely in the areas where transit also performs well, and vice versa. We do not conclude that the positive correlation confirms a complementary relationship over a competitive one between RH and transit. This is because the analysis is based on spatially aggregated data rather than single trip data revealing replacement or complementary travel. Rather, we suggest that future research focus on collecting a representative dataset of transit and ridehailing users to investigate trip-specific mode substitution and induced travel.

On the whole, despite pooled rides serving a larger range of communities and more peripheral areas as discussed above, we cannot find any statistical evidence that pooling compensates for transit deserts. What is more, pooling seems to offer less gap-filling than solo rides in areas where transit is poor, despite being more affordable. We speculate that ride-pooling might not be feasible or considered safe in transit-deserts.

## 4. Practical implications and suggested research

The findings suggest a number of implications for practice, RH operators and researchers. On the public policy side, the finding that ride-pooling demand correlates with vulnerable socio-economic living conditions measured by the SDI suggests that users in underserved community areas are benefitting from the convenience of an emerging mobility platform without paying the premium for solo rides. In terms of policy, this points to a need for greater focus dedicated to the positive socio-economic outcomes that TNCs can facilitate via pooled RH. By promoting ride-pooling, there are not only potential benefits from reduced congestion but also from users in disadvantaged community areas accessing more opportunities for employment and recreation. Thereby, public agencies ought to carefully differentiate RH taxes and regulations according to the type of service model, along with user-segment and locations, to avoid reducing mobility and accessibility for underserved communities.

Concerning the operational and business perspective, an important challenge arises when considering the greater spatial spread of pooled ride requests. Notably, to maintain effective shared on-demand service operations it is necessary to match multiple requester trajectories in real time. However, with only one in five riders requesting sharing, and the requests being geographically dispersed, it is challenging to efficiently tie together trajectories. At the same time, on the side of riders, to maintain a growing customer base and loyalty to pooling, it is important to ensure service quality. Research suggests that riders likely care more about trip time/cost than sharing itself (Lavieri & Bhat 2017). Therefore, understanding user expectations, and the socio-spatial context is necessary to promote demand for pooled services, to in turn enable more stream-lined matching and unlock the critical mass of pooling. Given the benefits to vulnerable community areas, RH operators and policy/mobility agencies have a strong motivation to work together to increase ride-pooling ridership.

On the research side there are three main take-aways. *First*, findings highlight the importance of studying contextual variables, such as socio-economic measures, more carefully. This calls for more research to disentangle how different mobility service offerings from the RH portfolio serves and affects different user segments and community areas. *Second*, RH service model effects are not monolithic. Specifically, the results point to a difference in magnitude or even in direction of explanatory effects when looking at different RH service models. *Third*, methodologically, this research uses a factor analysis-based index to study the overlapping factors of disadvantage that frequently affect communities. This helps overcome the underlying correlation between factors such as wealth, employment, and car-ownership, that jointly affect mobility decisions. An avenue for further work is to continue refining the indices that account for bundled factors to more accurately appraise the role of emerging mobility.

## 5. Conclusion

Innovative mobility services can be important tools to limit rising urban congestion and improve mobility for vulnerable populations. Yet, despite the significant growth in both the ridership and research on RH in recent years, findings on disparities in use have persisted not just along demographic dimensions such as income, gender, race/ethnicity, but also geographically. There is still limited understanding of the diverse demand patterns and the impact of varied services offered by RH operators (solo, pooled, shuttles, curb-to-curb, etc.). The goal of this study is to investigate the demand for RH services, focusing on the distinct socio-spatial patterns of solo requests versus ride-pooling. The analysis sheds light on how different emerging mobility services, with different sustainability, accessibility and equity implications, are used by diverse communities. We use a Spatial Durbin Model including measures of Social Disadvantage and Transit Accessibility applied to a publicly available dataset with 127 million RH records from the city of Chicago. The results show that density and vibrancy variables related to concentration of restaurants, population and younger residents, have similar effects on the demand for solo versus pooled rides. On the other hand, our analysis uncovers that pooling requests are geographically more dispersed and socially distinct from exclusive RH use. With regard to the three research questions posed in this work there are several implications.

- For Q1 we uncover that ride-pooling is utilized among a broader range of community areas outside the central business district, thereby serving more diverse communities. Comparing the solo and pooled ride determinants, we reveal that differences are mainly linked to community disadvantage. This suggests a novel connection between emerging mobility and vulnerable living conditions where pooled services can serve entirely different needs and populations than what has been observed in the research focused on solo RH. This has two important implications. *One*, for the spatial modeling of RH, disadvantages explain differences in demand, and also looms larger, that is, casts spillover effects across community areas. *Two*, the diffusion of pooling in underserved communities suggests an important socio-spatial dimension to consider in future work. Three, the more distributed demand pattern of pooled rides is tied to the sustainability of operations as critical demand-thresholds are harder to reach.

- For Q2 we develop an index that accounts for the bundled nature of socio-economic disadvantage. The SDI represents the only flipped sign in our spatial model: higher disadvantage is associated with more ride-pooling, and less demand for solo rides. Two implications arise. *One*, methodologically, there is value in using an index to account for highly correlated factors that affect RH demand. *Two*, a deeper analysis of the opportunity and barriers to accessing different RH models is needed. Analyzing service attributes, socio-economic circumstances and mobility context variables jointly is needed help understand which communities can access and benefit from pooling, and how it is used in practice.

- For Q3 we examined the impact of transit accessibility, finding that better rail transit access is correlated with more RH pickups (both solo and pooled). The findings call for more investigation to clarify why ride-pooling demand, seemingly a closer transit substitute, surges in transit-rich areas, then tapers of more rapidly in transit-underservedr community areas.

We note some important *caveats* of this study. First, owing to data censoring we are unable to distinguish Uber, Lyft and Via rides, leaving the different character and promotional strategies as unknown factors in shaping demand for solo versus pooled rides. Second, our trip data are not directly tied to rider sociodemographics. These are matched indirectly though the community area attributes and trip origin locations. Without precise rider data associated with each trip, it remains unknown whether the trips in high SDI areas, for example, are effectively requested by higher income trip-makers living in a disadvantaged community. Third, the data do not include information on drivers search/driving patterns or on operator locational/pricing algorithms which could affect the choice to use RH given that potential customers can view estimated waiting times and prices.

*Future research* should focus on further characterizing the differences between solo and pooled demand patterns (such as focusing on other variables such as trip length, timing, and duration), and analyzing their complex relationship with transit (buses and rail). There are potential benefits from reducing (solo) vehicle miles, and improving social outcomes, with increased use of pooling. Carefully designed stated and revealed preference/intercept surveys are needed to more fully capture the barriers to increased adoption of pooled rides.

Finally, with an eye to the future, while the RH industry tends to spearhead new forms of ride-sharing, currently and in the near future, societal values around sharing are changing drastically. As the world contends with the ongoing COVID-19 pandemic, and in many cases suspension of pooled RH services, it behooves researchers, policy-makers, and the RH industry to investigate the perceived risks of vehicle sharing, and the tolerance for returning to various forms of shared mobility.

## AUTHORS CONTRIBUTION

All authors contributed to all aspects of the study from model development, data processing, to analysis and interpretation of results, and manuscript preparation. All authors reviewed the results and approved the submission of the manuscript.

## COMPETING INTERESTS

We declare that we do not have competing interests.

**ACKNOWLEDGEMENTS**
The authors thank two JTG reviewers for detailed comments that helped improve this manuscript.